\documentclass[aps,prl,floatfix,twocolumn,showpacs,10pt]{revtex4-2}

\usepackage{graphicx}
\usepackage{dcolumn}
\usepackage{bm}
\usepackage{amssymb}
\usepackage{rotating} 
\usepackage{xcolor}

\begin{document}

\title{Quantum Simulation of Molecules without Fermionic Encoding of the Wave Function}

\author{David A. Mazziotti, Scott E. Smart, and Alexander R. Mazziotti}

\email{damazz@uchicago.edu}
\affiliation{Department of Chemistry and The James Franck Institute, The University of Chicago, Chicago, IL 60637}%

\date{Submitted January 27, 2021; Revised June 23, 2021}


\begin{abstract}
Molecular simulations generally require fermionic encoding in which fermion statistics are encoded into the qubit representation of the wave function.  Recent calculations suggest that fermionic encoding of the wave function can be bypassed, leading to more efficient quantum computations.  Here we show that the energy can be expressed as a functional of the two-electron reduced density matrix (2-RDM) where the 2-RDM is a unique functional of the unencoded $N$-qubit-particle wave function.  Contrasts are made with current hardware-efficient methods.   An application to computing the ground-state energy and 2-RDM of H$_{4}$ is presented.
\end{abstract}


\maketitle

{\em Introduction:} Quantum computers have the potential to revolutionize the computational sciences through an exponential advantage in certain classes of computations over their classical counterparts~\cite{Arute2019, Lloyd1996, Feynman1982}.  One of the most promising areas of application is the molecular sciences, especially molecular simulations of strongly correlated matter in which the electronic energies and properties of strongly correlated molecules are determined~\cite{Head-Marsden2020, McArdle2020, Sager2020, Arute2020, Bian2019, Paesani2017, OMalley2016, Wang2015, Nielsen2010, Du2010, Lanyon2010, Aspuru-Guzik2005, Bravyi2002, Abrams1997}.  A molecule or material is strongly correlated when a substantial number of orbitals are statistically dependent or in other words entangled~\cite{Kais2007}.  While the complexity of strongly correlated electronic calculations grows exponentially with the number of electrons on classical computers in the most difficult cases, the same calculations can in principle be performed without exponential cost on quantum computers.   Many-electron computations on quantum computers generally encode the particle statistics of electrons which are fermions into quantum bits or qubits in a process known as fermionic encoding~\cite{Jordan1928,Bravyi2002}.  While such encoding increases the complexity of the calculations, recent calculations suggest that fermionic encoding of the wave function can be bypassed for potentially more efficient quantum computations~\cite{Xia2020, Yordanov2020, Tang2019, Ryabinkin2020, Ryabinkin2018}.

In this Letter we examine molecular simulations without fermionic encoding of the wave function within the framework of reduced density matrix theory~\cite{Mazziotti2012b, Mazziotti2007, Coleman2000, Erdahl1978, Garrod1964, Coleman1963}.  Because electrons are indistinguishable with pairwise interactions, the energy can be expressed as a functional of the two-electron reduced density matrix (2-RDM)~\cite{Mazziotti2020b, Piris2017, Mazziotti2016, Verstichel2012, Mazziotti2011, Shenvi2010, Erdahl2007, Zhao2004, Mazziotti2004a, Mazziotti2002b, Nakata2001}.  In contrast to the classical theory, the quantum version of the variational 2-RDM theory measures the 2-RDM directly from $N$-electron wave function with potentially non-exponential scaling in $N$.  It generally requires fermionic encoding for both the wave function and the 2-RDM~\cite{Smart2019,Smart2020a,Smart2020} but not the Hamiltonian as in the conventional formulation of the variational quantum eigenvalue solver~\cite{Wei2020, Gard2020, Smart2020a, Kandala2019, McArdle2019, Kandala2017, McClean2016, Wecker2015, Peruzzo2014}.  Here we show that the 2-RDM can be expressed a unique functional of the unencoded $N$-qubit-particle wave function.  Significantly, the mapping preserves the entanglement complexity of the wave functions, and hence, the preparation of either wave function type should have a similar classical computational complexity, leading to potentially greater efficiency for the unencoded wave functions on a quantum computer.  To demonstrate the uniqueness of the functional, we prove that there exists an isomorphic mapping between the sets of $N$-fermion wave functions and $N$-qubit-particle wave functions.  Contrasts are made with current hardware-efficient methods~\cite{Kandala2017, McClean2016, McClean2017,McClean2018} as well as recent qubit-particle calculations~\cite{Xia2020, Yordanov2020, Tang2019, Ryabinkin2020, Ryabinkin2018}.  We also present an application of the theory to computing the ground-state energy and 2-RDM of H$_{4}$.

{\em Theory:} Each spin orbital in the calculation is represented by a qubit with the $|1 \rangle$ and $|0 \rangle$ states indicating that the orbitals are occupied and unoccupied, respectively~\cite{Nielsen2010,Abrams1997}.  An $N$-electron calculation with $r$ spin orbitals will have $r$ qubits with $N$ excited qubits in each configuration.  The $N$ excited qubits can be considered $N$ paraparticles~\cite{Green1953,Greenberg1965,Stolt1970,Wu2002},  particles with paranormal statistics in contrast to the normal statistics of fermions or bosons. For the $N$ qubit paraparticles to represent fermions,  the occupation of each orbital, however, must be complemented by information about the requisite anti-symmetry of the fermions, a process known as fermionic encoding of the qubits~ \cite{Jordan1928,Bravyi2002}.

In the variational 2-RDM calculation of the energy on a quantum computer, the wave function is prepared and its 2-RDM is measured at non-exponential cost. Both the preparation and measurement phases typically use fermionic encoding of the wave function. Once the 2-RDM is determined, that energy can be computed on the classical computer by.~\cite{Mazziotti2020b, Piris2017, Mazziotti2016, Verstichel2012, Mazziotti2011, Shenvi2010, Erdahl2007, Zhao2004, Mazziotti2004a, Mazziotti2002b, Nakata2001}
\begin{equation}
\label{eq:kd}
E = {\rm Tr}(^{2} K \; ^{2} D)
\end{equation}
where $^{2} K$ is the reduced Hamiltonian matrix, representing the reduced Hamiltonian operator
\begin{equation}
^{2} {\hat K} = \frac{N}{2} {\left ({\hat h}(1) + {\hat h}(2) \right )} + \frac{N(N-1)}{2} {\hat u}(12)
\end{equation}
in which ${\hat h}(1)$ denotes the kinetic and nuclear-attraction energies of electron~1 and ${\hat u}$(12) denotes the electron-electron repulsion, and $^{2} D$ is the 2-RDM.  Additional details can be found in Refs.~\cite{Mazziotti2011, Zhao2004, Mazziotti2004a, Mazziotti2002b, Nakata2001}  On an ideal quantum computer with zero noise and unlimited statistical sampling, the 2-RDM will be pure-state $N$-representable, meaning that it derives from at least one pure-state density matrix~\cite{Mazziotti2016a, Tennie2016, Benavides-Riveros2015, Chakraborty2014, Schilling2013, Altunbulak2008}.  Because current quantum computers have several sources of noise, however, the measured 2-RDM may require corrections in the form of error mitigation which can include the use of pure-state and ensemble $N$-representability conditions~\cite{Mazziotti2012b,Mazziotti2007,Coleman2000,Erdahl1978,Garrod1964,Coleman1963}.

Here we show that the non-degenerate ground-state fermionic 2-RDM can be written as a unique functional of the $N$-qubit-particle wave function.  To establish uniqueness, that the functional is bijective (one-to-one and onto), we prove that there is an isomorphic mapping between sets of fermionic wave functions and qubit-particle wave functions.  Importantly, the mapping is not only isomorphic (bijective while preserving the structure of the vector space) but also particle-number conserving.    Consider the wave function for $N$ fermions in $r$ orbitals
\begin{equation}
\label{eq:psif}
\psi_{F}(1,2,...,N) = \sum_{j_{1}<...<j_{N}}{c^{F}_{j_{1}j_{2}...j_{N}} B^{F}_{j_{1}j_{2}...j_{N}} }
\end{equation}
where $c^{F}_{j_{1}j_{2}...j_{3}}$ are expansion coefficients, $B^{F}_{j_{1}j_{2}...j_{N}}$ are the $N$-fermion basis functions
\begin{equation}
B^{F}_{j_{1}j_{2}...j_{N}} =  \sqrt{N!} \phi_{j_{1}}(1) \wedge \phi_{j_{2}}(2) \wedge ... \wedge \phi_{j_{N}}(N) ,
\end{equation}
$\phi_{j}$ is the $j^{\rm th}$ orbital, roman numbers denote the spatial and spin coordinates of the particles, and the $\wedge$ indicates the Grassmann wedge product~\cite{Slebodzinski1970, Mazziotti1998}.  The wedge product, which is rigorously defined in Appendix~A of Ref.~\cite{Mazziotti1998}, antisymmetrizes the product of orbitals with respect to the particle numbers, by adding (subtracting) the even (odd) permutations of the $N$ particles and dividing by the square root of the total number of permutations $N!$. The indices in the sums are restricted to range from lowest to highest, that is $j_{1} < j_{2} < ... < j_{N}$. The indices cannot be equal; otherwise, the wedge product vanishes in accord with the Pauli exclusion principle~\cite{Pauli1925} in which the orbital occupations are constrained to lie between zero and one.

Second, consider the wave function for $N$ qubit particles in $r$ qubits. The qubit particles can be understood as paraparticles because they obey neither fermionic nor bosonic statistics.  Like fermions they are restricted to no more than one particle per site, but like bosons they are symmetric in the exchange of two particles
\begin{equation}
\psi_{Q}(1,2,...,N) = \sum_{j_{1}<...<j_{N}}{c^{Q}_{j_{1}j_{2}...j_{N}} B^{Q}_{j_{1}j_{2}...j_{N}}}
\end{equation}
where $B^{Q}_{j_{1}j_{2}...j_{N}}$ are the $N$-qubit-particle basis functions
\begin{equation}
B^{Q}_{j_{1}j_{2}...j_{N}} =  \sqrt{N!} \phi_{j_{1}}(1) \vee \phi_{j_{2}}(2) \vee ... \vee \phi_{j_{N}}(N) .
\end{equation}
The qubit wave function $\psi_{Q}(1,2,...,N)$ is identical to the fermionic wave function $\psi_{F}(1,2,...,N)$ except for the replacement of the antisymmetric (Grassmann) wedge products $\wedge$ with the symmetric wedge products $\vee$~\cite{Slebodzinski1970, Mazziotti1998}.  The symmetric wedge products symmetrize the product of orbitals with respect to the particle numbers by adding all permutations of the $N$ particles and dividing the terms in the sum by $N!$.

Importantly, because the number of qubit particles is restricted by the qubit statistics to one per orbital like fermions, the expansion coefficients of both $\psi_{Q}(1,2,...,N)$ and $\psi_{F}(1,2,...,N)$---$c^{Q}_{j_{1}j_{2}...j_{N}}$ and $c^{F}_{j_{1}j_{2}...j_{N}}$---span exactly the same vector space of complex numbers of dimension $r$ choose $N$.    Therefore, there is an isomorphic mapping between the Hilbert space of $N$-fermion wave functions and the Hilbert space of $N$-qubit-particle wave functions.  Moreover, because of the folded indices ($j_{1} < j_{2} < ... < j_{N}$), the expansion coefficients, $c^{Q}_{j_{1}j_{2}...j_{N}}$ and $c^{F}_{j_{1}j_{2}...j_{N}}$, are also independent of the particle statistics and hence, interchangeable.  It follows that we can select the coefficients $c^{F}_{j_{1}j_{2}...j_{N}}$ from the set of possible qubit-particle coefficients $\{c^{Q}_{j_{1}j_{2}...j_{N}}\}$ without any approximation. Therefore, the wave function of $N$ fermion can be written as a functional of the wave function of $N$ qubit particles
\begin{equation}
\psi_{F}[\psi_{Q}](1,2,...,N) = \sum_{j_{1}<...<j_{N}}{c^{Q}_{j_{1}j_{2}...j_{N}} B^{F}_{j_{1}j_{2}...j_{N}}}
\end{equation}
which is identical to the fermion wave function in Eq.~(\ref{eq:psif}) except for the replacement of the fermion coefficients $c^{F}_{j_{1}j_{2}...j_{N}}$ by the qubit coefficients $c^{Q}_{j_{1}j_{2}...j_{N}}$. The antisymmetrized basis functions $B^{F}_{j_{1}j_{2}...j_{N}}$ cause the qubit coefficients $c^{Q}_{j_{1}j_{2}...j_{N}}$ to be interpreted as fermions.  Integration of the density matrix associated with the $\psi_{F}[\psi_{Q}](1,2,..,N)$ over the coordinates of all fermions except 1 and 2 yields the fermionic $N$-representable 2-RDM
\begin{equation}
^{2} D[\psi_{Q}]^{12}_{{\bar 1}{\bar 2}} = \int{\psi_{F}[\psi_{Q}](1...N) \psi^{*}_{F}[\psi_{Q}]({\bar 1}...N) d3..dN}
\end{equation}
as a functional of the $N$-qubit-particle wave function.  It is known by Rosina's theorem~\cite{Mazziotti1998} that the $N$-electron wave function of a non-degenerate $N$-electron ground state is a bijective functional of the  $N$-representable 2-RDM.  Therefore, for a non-degenerate $N$-electron ground state the functional connecting the $N$-representable 2-RDM and the $N$-qubit-particle wave function is also bijective.

The above argument establishes the result, but practical application on a quantum computer requires us to develop the result in second quantization.  In first quantization the symmetry of the particles---bosons, fermions, or paraparticles---is contained in the wave function, but in second quantization the symmetry of the particles is contained in the second-quantized operators and the anti-commutation (or commutation) relations that govern them~\cite{Berezin1966}.  Hence, the state of the wave function in second quantization is agnostic to the particle statistics
\begin{equation}
| \psi \rangle = \sum_{j_{1}<j_{2}<...<j_{N}}{c_{j_{1}j_{2}...j_{N}} \phi_{j_{1}} \phi_{j_{2}}  ... \phi_{j_{N}}}
\end{equation}
where the $c_{j_{1}j_{2}...j_{N}}$  are the same expansion coefficients as in first quantization while each of the $N$-particle basis functions $| \phi_{j_{1}} \phi_{j_{2}} ... \phi_{j_{N}} \rangle$ denotes an ordered product of orbitals that is independent of the particle statistics.  For convenience, we introduce fermion and qubit-particle labeled wave functions to indicate how the wave functions are prepared. The fermion wave function
\begin{equation}
| \psi_{F} \rangle = \sum_{j_{1}<j_{2}<...<j_{N}}{c^{F}_{j_{1}j_{2}...j_{N}} \phi_{j_{1}} \phi_{j_{2}}  ... \phi_{j_{N}}}
\end{equation}
is prepared with fermion second-quantized operators and the qubit-particle wave function
\begin{equation}
| \psi_{Q} \rangle = \sum_{j_{1}<j_{2}<...<j_{N}}{c^{Q}_{j_{1}j_{2}...j_{N}} \phi_{j_{1}} \phi_{j_{2}}  ... \phi_{j_{N}}}
\end{equation}
is prepared with qubit second-quantized operators.

Typically, on a quantum computer the 2-RDM is computed from a wave function prepared with fermion second-quantized operators~\cite{Smart2020}
\begin{equation}
^{2} D^{pq}_{st} = \langle \psi_{F} | {\hat a}^{\dagger}_{p}   {\hat a}^{\dagger}_{q}  {\hat a}_{t}  {\hat a}_{s} | \psi_{F} \rangle
\end{equation}
where ${\hat a}^{\dagger}_{p}$ and ${\hat a}_{p}$ are the creation and annihilation operators for a fermion in orbital $p$.  However, because the coefficients in the fermion and qubit-particle wave functions --- $c^{Q}_{j_{1}j_{2}...j_{N}}$ and $c^{F}_{j_{1}j_{2}...j_{N}}$ --- span the same vector space of complex numbers of dimension $r$ choose $N$ and are identical to their definitions in first quantization, we can uncouple the particle statistics of the preparation from the measurement and express the fermionic 2-RDM in terms of the qubit-particle wave function $\psi_{Q}(1,2,..,N)$  without any approximation
\begin{equation}
\label{eq:dq}
^{2} D^{pq}_{st} = \langle \psi_{Q} | {\hat a}^{\dagger}_{p}   {\hat a}^{\dagger}_{q}  {\hat a}_{t}  {\hat a}_{s} | \psi_{Q} \rangle .
\end{equation}
The second-quantized operators with their fermionic anti-commutation relations cause the qubit wave function $ | \psi_{Q} \rangle$ and its coefficients $c^{Q}_{j_{1}j_{2}...j_{N}}$ to be interpreted as fermionic in generation of the 2-RDM.  As in first quantization, we have demonstrated that the fermionic pure-state $N$-representable 2-RDM is expressible as a unique functional of the $N$-qubit-particle wave function.  Again by Rosina's theorem~\cite{Mazziotti1998} the functional is bijective for non-degenerate ground states.  Both the first and second-quantized results are immediately generalizable to show that  the fermionic pure-state $N$-representable $p$-RDM is expressible as a unique functional of the $N$-qubit-particle wave function for any $p \le N$.

To implement the measurement of the 2-RDM elements in Eq.~(\ref{eq:dq}) on a quantum computer, we express the fermionic operators in terms of qubit-particle operators through a Klein transformation~\cite{Araki1961, Greenberg1977, Klein1938} known as the Jordan-Wigner transformation~\cite{Jordan1928}
\begin{eqnarray}
{\hat a}^{\dagger}_{p} & = & {\hat \sigma}^{\dagger}_{p} {\hat \chi}^{\dagger}_{p} \\
{\hat a}_{p} & = & {\hat \chi}_{p} {\hat \sigma}_{p}
\end{eqnarray}
to obtain
\begin{equation}
^{2} D^{pq}_{st} = \langle \psi_{Q} | {\hat \sigma}^{\dagger}_{p}  {\hat \chi}^{\dagger}_{p}  {\hat \sigma}^{\dagger}_{q}  {\hat \chi}^{\dagger}_{q} {\hat \chi}_{t}  {\hat \sigma}_{t} {\hat \chi}_{s} {\hat \sigma}_{s} | \psi_{Q} \rangle
\end{equation}
where the ${\hat \sigma}^{\dagger}_{p}$ and ${\hat \sigma}_{p}$ are the creation and annihilation operators for a qubit particle in orbital $p$ and the unitary Klein operator is
\begin{equation}
{\hat \chi}_{p} = e^{i \pi \sum_{q=1}^{p-1}{{\hat \sigma}^{\dagger}_{q} {\hat \sigma}_{q}}} .
\end{equation}
The operator ${\hat \chi_{p}}$ introduces the necessary phase changes to convert the commutation relations of qubits into the anti-commutation relations of fermions~\cite{Araki1961, Greenberg1977, Klein1938}. The use of a Klein transformation to make qubit second-quantized operators represent fermion second-quantized operators is known as fermionic encoding of the qubits. By decoupling of the particle statistics of the preparation from the measurement, we can prepare the wave function with qubit-particle operators and measure the 2-RDM with fermionic-encoding qubit operators.  Hence, we can replace the computationally more expensive fermionic encoding of the wave function with the fermionic encoding of just the 2-RDM.   Use of the encoding 2-RDM in Eq.~(\ref{eq:kd}) to compute the energy as in the variational 2-RDM method~\cite{Mazziotti2020b, Piris2017, Mazziotti2016, Verstichel2012, Mazziotti2011, Shenvi2010, Erdahl2007, Zhao2004, Mazziotti2004a, Mazziotti2002b, Nakata2001} is similar to the encoding of the Hamiltonian in the hardware-efficient methods~\cite{Kandala2017, McClean2016, McClean2017}.  However, unlike the hardware-efficient methods where the implicit 2-RDM may not be $N$-representable due to a lack of particle-number conservation, the $N$-qubit-particle wave function formally generates an $N$-representable 2-RDM.

{\em Results:} The mean-field wave functions can be readily prepared by applying $N$ qubit-particle creation operators to excite the qubits corresponding to the $N$ energetically lowest orbitals~\cite{Jiang2018}.  Correlated qubit-particle wave functions can be generated by applying two-or-more-qubit-particle unitary transformations to the mean-field reference wave function by an adiabatic construction~\cite{Babbush2014}, an imaginary-time evolution~\cite{McArdle2019,Motta2019}, or a variational quantum eigensolver (VQE)~\cite{Wei2020, Gard2020, Smart2020a, Kandala2019, McArdle2019, Kandala2017, McClean2016, Wecker2015, Peruzzo2014}.  Based on the present result, in a VQE the $p$-body anti-Hermitian fermionic operators in the UCC can be replaced by $p$-body anti-Hermitian qubit-particle excitation operators without fermionic encoding. This replacement is computationally significant because fermionic encoding causes a $p$-body operator to become an $r$-body operator whose exponentiation yields an $O(r)$ number of two-qubit CNOT gates where $r$ is the rank of the orbital basis set such that $r \gg p$ (in contrast, exponentiation of the $p$-qubit-body operator generates a circuit of CNOT gates that scales independently of system size $O(1)$).  Recent results by Xia and Kais~\cite{Xia2020} and Ryabinkin {\em et al.}~\cite{Ryabinkin2020, Ryabinkin2018} show that accurate results can be obtained from a qubit-particle UCC ansatz at a reduced computational cost.

Two general classes of fermionic unitary transformations include: (i) the unitary coupled cluster (UCC) ansatz~\cite{Xu2020, Shen2017, Bartlett2007} and (ii) the anti-Hermitian contracted Schr{\"o}dinger equation (ACSE) ansatz~\cite{Smart2020, Mazziotti2007a, Mazziotti2007d, Mazziotti2006, Mazziotti1998}.  While the degree of excitations must equal $N$ to become exact in the UCC ansatz, in the ACSE ansatz the wave function is expressed as a product of general two-body unitary transformations (see Ref.~\cite{Mazziotti2007a} for details).  It has been shown that the product of general two-body unitary transformations is an exact ansatz for the $N$-electron wave function~\cite{Evangelista2019, Mazziotti2020, Mazziotti2004}.  As in the UCC ansatz, the exponentiation of general two-body anti-Hermitian fermionic operators in the ACSE ansatz~\cite{Smart2020} can be replaced by the exponentiation of general two-body anti-Hermitian qubit-particle operators with potentially significant computational savings.  Qubit excitations have recently been implemented in  an ACSE-related theory, known as ADAPT-VQE, on a quantum simulator for small molecules~\cite{Yordanov2020} with similar accuracy as the fermionic methods.

Here we compute the ground-state energy and 2-RDM of H$_{4}$ by a quantum solution of the ACSE presented in Ref.~\cite{Smart2020}.  The geometry of H$_{4}$ is taken to be a linear chain with equally spaced bonds of 0.88821~\AA.  We use a minimal Slater-type orbital with 3 Gaussian functions (STO-3G)~\cite{Hehre1969}.  The simulation is performed without stochastic error.  Each iteration of the ACSE applies a two-body unitary transformation based on the residual of the ACSE.   We compare the accuracy and computational cost of using fermionic and qubit-particle wave functions in Fig.~1.   The log-base-10 error in the ground-state energy of H$_{4}$, relative to the energy from full configuration interaction, reveals nearly identical convergence---decreases in energy with iteration number---of the fermionic and qubit-particle wave functions.  We also show in Fig.~1 that the cumulative number of CNOT gates increases more slowly for the qubit-particle wave function than the fermionic wave function.  The lower qubit-particle CNOT count results from the omission of the exponentiated Klein operators.  Further details and examples will be presented elsewhere.

\begin{figure}[htp!]

\includegraphics[scale=0.50]{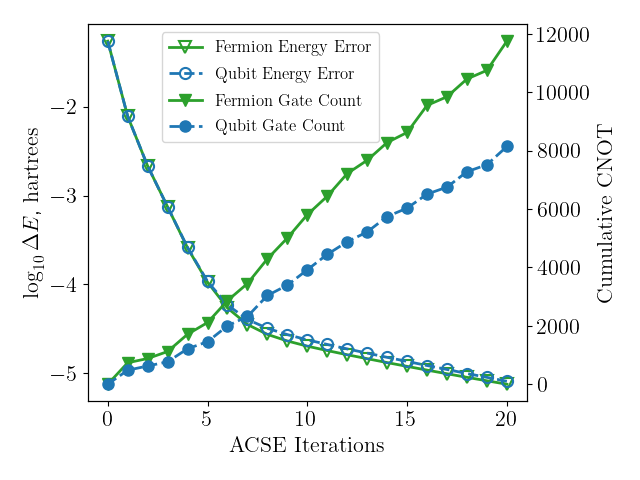}

\caption{The accuracy and computational cost of using fermionic and qubit-particle wave functions are compared. The log-base-10 error in the ground-state energy of H$_{4}$ reveals nearly identical convergence for both types of wave functions.  Moreover, the cumulative number of CNOT gates increases more slowly for the qubit-particle wave function.}

\label{fig:acse}

\end{figure}

{\em Discussion and Conclusions:} While a 2-RDM can always be parameterized by its wave function to be $N$-representable, the $N$-representability conditions are typically formulated in terms of the 2-RDM to avoid the exponential computational scaling of the $N$-electron wave function~ \cite{Mazziotti2012b, Mazziotti2007, Coleman2000, Erdahl1978, Garrod1964, Coleman1963}.  With a quantum computer, however, the $N$-fermion wave function can be prepared and the 2-RDM measured at non-exponential cost, and hence, the 2-RDM can be kept $N$-representable by its measurement from its fermionic wave function. Here we have shown that the set of fermionic 2-RDMs can be expressed as a unique functional of the $N$-qubit-particle wave function and hence, the requirement that the prepared wave function be fermionic can be relaxed.  The functional shows that the fermionic encoding of the wave function is not necessary for generating a strictly $N$-representable 2-RDM as long as the encoding is employed in the measurement of the 2-RDM.  Previous work has recognized the $N$-representability conditions as an important resource for the mitigation of noise-related errors~\cite{Smart2019, Rubin2018, Foley2012}.  The present result emphasizes that the $N$-representability of the 2-RDM has a central role in quantum computing.

Variational algorithms for molecular simulation on quantum computers prepare the wave function and measure the fermionic 2-RDM.  With knowledge of the 2-RDM the energy can be inexpensively computed on the classical computer by Eq.~(\ref{eq:kd}).  Both the wave function preparation and the 2-RDM measurement typically use fermionic encoding of the qubits in which the qubit-particle operators are transformed to restore fermionic statistics~\cite{Jordan1928,Bravyi2002}.  Here we have shown that there is an isomorphic mapping between $N$-fermion and $N$-qubit-particle wave functions, allowing us to express the 2-RDM as a functional of the $N$-qubit-particle wave function with the functional being unique (bijective) for a non-degenerate ground state.  Because the mapping preserves the entanglement complexity of the wave functions, both wave function types should have a similar classical complexity with the qubit-particle wave functions being potentially more efficient on qubit-based quantum computers.  In contrast, the hardware-efficient algorithms~\cite{Kandala2017, McClean2016, McClean2017} have a many-to-one mapping between the generic qubit wave functions and an implicit 2-RDM with the generic qubit wave functions displaying fermionic, bosonic, and mixed-particle statistics.  Such a many-to-one mapping as well as a lack of particle-number conservation can lead to observed optimization difficulties due to over-parametrization such as barren plateaus~\cite{McClean2018}.  The theory also provides an explanation for the promising results recently obtained by Xia and Kais~\cite{Xia2020}, Yordanov {\em et al.}~\cite{Yordanov2020}, Tang {\em et al.}~\cite{Tang2019}, and Ryabinkin {\em et al.}~\cite{Ryabinkin2020, Ryabinkin2018} on quantum simulators from the use of qubit-excitation wave functions.  While additional research is needed to assess the full potential of qubit-particle wave functions,  the recent and present results indicate that they represent a promising direction for more efficient molecular quantum simulations.

\begin{acknowledgments}

D.A.M. gratefully acknowledges the Department of Energy, Office of Basic Energy Sciences Grant DE-SC0019215 and the U.S. National Science Foundation Grants CHE-1565638 and No. CHE-2035876.

\end{acknowledgments}

\bibliography{QComputersV3,RDM2,Parastatistics}

\end{document}